\documentclass[conference]{IEEEtran}
\IEEEoverridecommandlockouts

\usepackage{cite}
\usepackage{amsmath,amssymb}
\usepackage{booktabs}
\usepackage{multirow}
\usepackage{graphicx}
\usepackage{url}
\usepackage{balance}
\usepackage{hyperref}

\newcommand{\tabstyle}{\footnotesize\setlength{\tabcolsep}{3pt}}
\newcommand{\hdrrow}{Dataset & Method & Macro-F1 & Cost \\}

\title{Stop Early, Stay Simple: Evidence Acquisition for Power System Fault and Attack Diagnosis}

\author{
    \IEEEauthorblockN{
        S M Zia Ur Rashid\textsuperscript{1,\dag}\thanks{\textsuperscript{\dag}Corresponding author.}, 
        Sazidul Islam Hira\textsuperscript{2}, 
        Rupesh Sah\textsuperscript{1}, and 
        Suman Rath\textsuperscript{1}
    }
    \IEEEauthorblockA{
        \textsuperscript{1}Department of Electrical and Computer Engineering, The University of Tulsa, Tulsa, OK, USA\\
        \textsuperscript{2}Department of Computer Science and Engineering, International Islamic University Chittagong, Chittagong, Bangladesh\\
        Email: \textsuperscript{1}\{ziaur-rashid, rks9560, suman-rath\}@utulsa.edu, \textsuperscript{2}sazidulislam.mail@gmail.com
    }
}

\begin{document}
\maketitle

\begin{abstract}
Diagnostic models for cyber-physical power systems usually inspect every telemetry source for every event, although some events are clear from one source and others need several. We ask when a learned policy for choosing that evidence is worth its added complexity. To answer this, we present ActiGrid, a benchmark framework for active evidence acquisition that pairs a classifier trained to tolerate missing modalities with a Double Deep Q-Network (DQN), which either requests another source or stops and diagnoses but never operates grid actuators. On the public TU Delft digital substation and MSU/ORNL power system attack benchmarks, we compare the DQN with fixed subsets and orders, calibrated entropy stopping, a greedy value of information rule, and tree ensembles trained per subset. The learned policy does adapt to what it observes, yet it is statistically indistinguishable from entropy stopping on TU Delft and Pareto dominated by simple controls on MSU/ORNL, while compact subset classifiers outperform it while reading fewer sources on both benchmarks. Entropy stopping run over those subset classifiers reaches the highest mean Macro-F1 on TU Delft (0.928 while reading 57\% of the sources) but brings no saving over a static three source subset on MSU/ORNL, where no method exceeds 0.45 Macro-F1. Selective acquisition can pay off, but on these benchmarks it does not need a learned policy.
\end{abstract}

\begin{IEEEkeywords}
Active feature acquisition, cyber-physical systems, deep reinforcement learning, fault and attack diagnosis, power systems security, sequential decision-making
\end{IEEEkeywords}

\section{Introduction}
Power system operators now monitor the grid through electrical measurements, protection signals, communication traffic, and host telemetry, a combination that improves observability but also widens the cyber-physical attack surface \cite{Sridhar2012}. False data injection shows how manipulated measurements can corrupt operational inference \cite{Liu2009}, and operational technology monitoring must meet reliability and safety requirements that differ from those of IT analytics \cite{NIST2023}.

Most diagnostic classifiers quietly assume that every configured source is available whenever a prediction is made, which is convenient but not always needed. Network visible injections may be recognizable without any physical or host signal, whereas hard cases that could be either a fault or an attack may need several independent views. Heterogeneous, time synchronized evidence already helps power system diagnosis \cite{Pan2015}, and digital substations add IEC~61850 GOOSE and sampled value (SV) traffic together with resource telemetry \cite{Mackiewicz2006,Cibin2026}. So a second question sits next to classification, namely which evidence to inspect before diagnosing.

Active sensing and active feature acquisition study this kind of cost aware observation \cite{Yu2009,Shim2018,Li2021,Valancius2024}. In this work we apply the idea to whole evidence modalities and use the term \emph{evidence acquisition} for the choice of which source to reveal next and when to stop. We present ActiGrid, a benchmark framework for active evidence acquisition in grid diagnosis, in which a reinforcement learning (RL) policy makes this choice over a classifier trained for missing modalities and is compared with fixed, rule based, and tree baselines, while the policy never controls breakers, relays, or protection logic. Every modality receives the same normalized cost, so the reported cost is the average fraction of modalities read and does not measure bandwidth, energy, latency, or operator effort.

We frame this work as a methodological benchmark rather than as a case for RL by default. Learning to acquire or stop is well established for classification with costly features, including deep RL formulations \cite{Janisch2019}. Our action spaces are small, though, with three modalities on TU Delft and five on MSU/ORNL, so exhaustive static subsets and simple sequential rules are realistic competitors rather than weak baselines. Our study centers on three questions about learned evidence acquisition. (Q1)~With one shared classifier, does the learned policy beat fixed and sequential rules? (Q2)~Does the answer hold on new splits and against classifiers trained for each subset? (Q3)~Does the policy really adapt, and how fragile is it to corrupted early evidence? In answering these questions, we make the following four contributions.
\begin{itemize}
\item We cast fault and attack diagnosis as sequential modality acquisition over one shared classifier, which isolates the acquisition decision.
\item We compare a Double DQN on that classifier with full fusion, every fixed subset and order, calibrated entropy stopping, and a greedy value of information rule, and we measure mask-wise calibration (Q1).
\item We retest the comparison on three group disjoint splits against tree ensembles trained per subset and against entropy stopping run over those trees (Q2).
\item We show how the learned policy branches after its first observation and how corrupting the first acquired source changes its accuracy and cost (Q3).
\end{itemize}
The rest of the paper is organized as follows. Section~\ref{sec:related} reviews related work, Section~\ref{sec:method} presents the ActiGrid framework and its baselines, Section~\ref{sec:design} describes the datasets and evaluation protocol, Section~\ref{sec:results} answers Q1 to Q3, Section~\ref{sec:discussion} discusses implications and limitations, and Section~\ref{sec:conclusion} concludes the paper.

\section{Related Work}
\label{sec:related}
Our study draws on three lines of work, namely fault and attack diagnosis in power systems, active feature acquisition, and learning with missing modalities and tabular data. Power system attack diagnosis has long combined physical and communication measurements to separate disturbances from malicious behavior. The MSU/ORNL data were released with machine learning baselines for discriminating disturbances from cyber attacks \cite{Hink2014}, and Pan \emph{et al.} classify disturbances and attacks from their heterogeneous, time synchronized measurements and logs \cite{Pan2015}. Abedi \emph{et al.} distinguish faults from cyber attacks in near real time by pairing statistical cyber anomaly detection with fault detection \cite{Abedi2023}, the TU Delft dataset exposes physical, resource, and packet level evidence \cite{Cibin2026}, and Presekal \emph{et al.} detect intrusions in substation traffic with semi-supervised learning and similarity clustering \cite{Presekal2026}. These works read every configured source for every event, which is the assumption we question here.

Outside power systems, active sensing asks whether all features must be obtained before prediction \cite{Yu2009}, and EDDI picks the next feature by its estimated information gain \cite{Ma2019}, the greedy principle that our Greedy-VOI control follows. Acquire or stop policies have also been learned with RL \cite{Shim2018,Janisch2019}, with generative surrogate models \cite{Li2021}, with learned approximations of greedy mutual information selection \cite{Covert2023}, and with nongreedy acquisition oracles \cite{Valancius2024}. Budgeted sequential classification pursues a similar goal with a cascade of classifiers over growing feature sets and learned rules for deferring to the next stage \cite{Trapeznikov2013}, and our sequential HGB control is a simple instance of that idea with entropy thresholds in place of learned rules. We claim no novelty for active acquisition, deep Q-learning \cite{Mnih2015}, or Double Q-learning \cite{Hasselt2016}, and our work differs from these strands in two respects, since we act on whole modalities rather than single features and ask whether a learned policy earns its complexity against strong fixed and sequential alternatives in this domain.

Training with randomly dropped modalities keeps a fusion network usable when sources are missing, an idea introduced as ModDrop for gesture recognition \cite{Neverova2016}, and temperature scaling is the standard post hoc correction for overconfident softmax outputs \cite{Guo2017}, which we fit separately for each acquisition mask. Classifier choice also matters because both benchmarks are tabular. Studies of tabular learning keep finding that gradient boosted trees are hard for neural models to beat \cite{Gorishniy2021,Grinsztajn2022}, so a fair comparison needs strong non-neural baselines, and we therefore pair the shared neural backbone with random forests, histogram gradient boosting, and XGBoost \cite{Chen2016} per subset.

\section{The ActiGrid Framework}
\label{sec:method}
\begin{figure}[t]
\centering
\includegraphics[width=0.9\columnwidth]{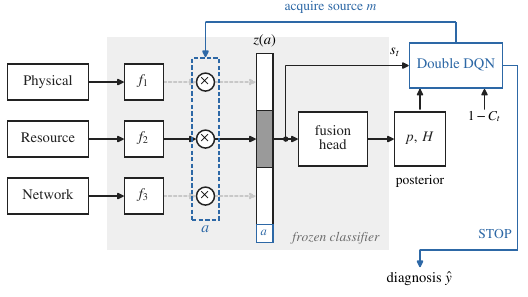}
\caption{ActiGrid acquisition loop, drawn for the three TU Delft sources.}
\label{fig:method}
\end{figure}

ActiGrid consists of a classifier trained to tolerate missing modalities, a learned acquisition policy that runs on top of it, and a set of simpler acquisition baselines. Fig.~\ref{fig:method} shows the acquisition loop, in which the frozen classifier scores whatever sources have been read so far and the policy decides whether to read another source or to stop and diagnose.

\subsection{Classifier for Missing Modalities}
Let $M$ be the number of evidence modalities, and let a source specific encoder map modality $m$ to a latent vector $z_m=f_m(x_m)$. For an acquisition mask $a\in\{0,1\}^{M}$, the classifier receives
\begin{equation}
 z(a)=[a_1z_1\Vert\cdots\Vert a_Mz_M\Vert a],
 \label{eq:mask}
\end{equation}
where $\Vert$ denotes concatenation. We append the mask itself so that a missing source is never confused with an observed one whose representation happens to lie near zero.

Each encoder has a 128-unit hidden layer and a 64-dimensional latent output with LayerNorm, GELU, and dropout of 0.2, so the input in \eqref{eq:mask} has $65M$ dimensions, which is 195 on TU Delft and 325 on MSU/ORNL. The fusion head maps it through layers of 192 and 128 units. We train with class weighted cross entropy and AdamW (learning rate $10^{-3}$, weight decay $10^{-4}$) on batches of 512, clip gradients at 5, and stop early with a patience of 7 over at most 35 epochs.

During training, each modality is dropped independently with probability 0.25. Another 2\% of samples see no modality at all, and any other empty mask gets one random modality back. This exposes the network to partial inputs, but not evenly, since single modality masks make up about 15.3\% of training masks on TU Delft and only 1.5\% on MSU/ORNL, against 41.3\% and 23.3\% for full masks. Low cost states are therefore rare during training, and dropout also does not make the probabilities well calibrated under every mask, two points to which Section~\ref{sec:results} returns.

\subsection{Sequential Acquisition Policy}
At step $t$ the policy sees $z(a)$ from \eqref{eq:mask}, which holds the masked encoder outputs and the mask itself, together with the class posterior, its entropy, and the cost headroom $1-C_t$, where $C_t$ is the cost incurred so far. For any fixed cost vector the headroom is determined by the mask and adds no information. We keep it so that the same state applies when budgets or costs vary between episodes. The state thus has $65M$ entries plus one per class and two scalars, which is 203 on TU Delft and 332 on MSU/ORNL. Actions are the unobserved modalities plus STOP, and invalid actions are masked. We freeze the classifier before policy training, and the policy never sees raw values of a modality it has not acquired.

The Double DQN has two ReLU layers of 256 units and is trained with AdamW at learning rate $2\times10^{-4}$, a replay buffer of 80,000 transitions, and batches of 256, using discount $\gamma=0.97$, the Smooth L1 loss, gradient clipping at 5, and a target network update every 500 optimization steps. Training runs for 20,000 class balanced episodes, and exploration decays linearly from $\epsilon=1.0$ to 0.05 over the first 70\% of them.

For an acquisition action, the reward is
\begin{equation}
 r_t=\operatorname{clip}(\log p_y^{t+1}-\log p_y^t,-2,2)-\lambda c_m,
 \label{eq:reward}
\end{equation}
where $y$ is the true class, $p_y^t$ is its softmax probability under the frozen classifier after step $t$, and $c_m$ is the cost of modality $m$. These probabilities are uncalibrated and floored at $10^{-8}$, and temperature scaling enters only the calibrated entropy baseline, never DQN training. The true class appears only inside the training reward. STOP returns $+2$ for a correct diagnosis and $-2$ otherwise, and confusing a fault with an attack, in either direction, costs a further $-1$. Without clipping and discounting, these terms would telescope to $\log p_y^T-\log p_y^0$, a difference of the potential $\Phi=\log p_y$, but because $\Phi$ is not reset at STOP and the terms are clipped and discounted, \eqref{eq:reward} lacks the policy preserving form of \cite{Ng1999} and acts as heuristic shaping toward a better diagnosis. Our original operating point was $\lambda=0.5$, and the hardening study also tries 0.25 and 0.75.

\subsection{Simple Acquisition Baselines}
Each of the five MSU/ORNL modalities costs 0.2 and each of the three TU Delft modalities costs $1/3$. On the shared backbone we compare the DQN with full fusion, every nonempty fixed subset, and every fixed acquisition order with entropy stopping. We also train random forest (RF), histogram gradient boosting (HGB), and XGBoost classifiers for each subset.

Two further sequential controls need no RL at all. \emph{Calibrated entropy stopping} fits one temperature per mask on validation data and only then picks a stopping threshold. \emph{Greedy-VOI} (value of information) predicts how much one more step with each remaining modality would reduce entropy and acquires the source with the largest predicted reduction per unit cost. Its predictor is one small HGB regressor per modality, fit on frozen backbone states from a sample of 7,000 training rows for MSU/ORNL and from all 3,246 training rows for TU Delft.

Both controls still run on the shared neural backbone, so to remove the difference in classifier family we also run fixed order entropy stopping directly over the per subset HGB models, which we call \emph{sequential HGB}. Each subset model gets one temperature fitted on one half of the validation set, while the order, the threshold, and the static subset we compare against are chosen on the other half. All three controls are deliberately simple, which suits an action space this small.

\section{Experimental Design}
\label{sec:design}
Two public benchmarks with group disjoint splits and two complementary protocols let us separate the effect of the acquisition policy from that of the classifier.

\subsection{Datasets and Splits}
\textbf{MSU/ORNL.} The processed benchmark has 78,377 samples from 15 data files with 128 predictors, grouped into four phasor measurement unit (PMU) and relay modalities of 29 features each and one cyber log modality of 12 features. Two of the five labels are non-malicious (normal, and physical fault or maintenance), and three are data injection, remote trip, and relay setting attacks. Holding out whole files (9/3/3) gives 47,471/15,272/15,634 samples for training/validation/test, and under this protocol no model we tried exceeds 0.45 Macro-F1, including tree ensembles on all 128 features. Because each file records the same 37 scenarios \cite{Hink2014,Pan2015} and data injection imitates valid faults by design \cite{Hink2014}, file holdout does not hold out scenario types, and the fault and injection labels are hard to separate. We have not isolated how much each factor limits the scores, so on this benchmark we compare methods only against one another.

\textbf{TU Delft.} Our processing pipeline yields 5,838 aligned one second windows from 79 usable runs, labeled with six classes that cover normal operation, physical faults, reconnaissance, resource exhaustion, and injection attacks on GOOSE and SV traffic. The modalities hold 40 physical RSCAD features, 60 resource monitoring features, and 16 network features from PCAP and PCAPNG traffic, and the primary split assigns 47/16/16 runs and 3,246/1,276/1,316 windows to training/validation/test. Frame level packet labels define the network attack windows, and the physical fault indicator serves only as ground truth. We remove PID, USER, and COMMAND identifiers, convert cumulative resource counters to rates, join resource data using past values only, and align the affected clocks with a label independent shift of $+7200$~s. We exclude two GOOSE baseline runs that could be paired with more than one resource file rather than silently pick one, and we label resource exhaustion recordings at the recording level because they carry no precise onset time.

\subsection{Evaluation Protocol}
\label{sec:protocol}
Macro-F1 is the primary metric because both datasets are imbalanced, and Tables~\ref{tab:controlled}, \ref{tab:hardening}, and~\ref{tab:hgbseq} report the mean $\pm$ standard deviation (SD) over seeds or partitions. In the \emph{shared backbone protocol}, every policy for a given seed uses the same trained fusion classifier, so any difference comes from the acquisition decision alone. Real deployments could instead train a classifier for the one subset they read, so we also report the stronger frontier of tree models fit per subset.

The controlled experiment trains ten seeds on one fixed group disjoint split, so its intervals describe training variation rather than independent test populations. To test split sensitivity, we build three group disjoint partitions with \emph{partition seeds} 42, 123, and 456, of which partition 42 is the primary split, and on each we train one fusion model and one DQN with model seed 4242 and choose RF, HGB, and XGBoost subsets by validation Macro-F1. Separately, three backbones on the primary split use \emph{initialization seeds} 42, 123, and 456, so that the $\lambda$ frontier varies the initialization while the data stay fixed. These backbones serve the $\lambda$ frontier, the non-RL controls, and every seed 42 analysis in Section~\ref{sec:results}, so the partition 42 model in Table~\ref{tab:hardening} shares its data with initialization seed 42 but not its weights.

For sequential policies we inspect the whole Pareto frontier of Macro-F1 against cost rather than a single operating point, and we measure mask-wise calibration by expected calibration error (ECE). On the primary split, a group bootstrap with 3,000 replicates gives intervals for the difference between DQN and entropy stopping, and we do not read a failure to detect a difference as evidence of equivalence.

All runs used Kaggle notebooks, with CPU only for the controlled and hardening runs, where one MSU/ORNL policy took about 18 minutes to train. The Greedy-VOI analysis reused frozen checkpoints on a Tesla T4 GPU and retrained no fusion, DQN, or tree model, and sequential HGB fits only the per subset trees and their temperatures.

\section{Results}
\label{sec:results}
On the shared backbone (Q1), the TU Delft DQN reaches Macro-F1 0.7917 at cost 0.703 (Table~\ref{tab:controlled}), keeping 97.0\% of the full fusion score with 29.7\% fewer modality units and far ahead of the best two source subset on the same classifier. Fixed order entropy stopping reaches 0.7868 at slightly lower cost, and the paired DQN minus entropy difference of $+0.0049$ (95\% confidence interval $[-0.0043,0.0141]$, Wilcoxon $p=0.432$) does not establish a DQN advantage. MSU/ORNL is less kind to learned acquisition, since the DQN reaches 0.3769 at cost 0.754 while a fixed three source subset reaches 0.4011 at cost 0.60, a difference of $-0.0242$ (95\% interval $[-0.0360,-0.0124]$), so even before stronger classifiers enter, a compact static design is the better choice. The answer to (Q1) is thus negative on MSU/ORNL and inconclusive on TU Delft.

\begin{table}[t]
\centering
\tabstyle
\caption{Shared backbone comparison on the primary split over ten training seeds.}
\label{tab:controlled}
\begin{tabular}{llcc}
\toprule
\hdrrow
\midrule
\multirow{4}{*}{TU Delft}
 & Full fusion & $\mathbf{0.8158\pm0.0323}$ & 1.000 \\
 & DQN ($\lambda=0.5$) & $0.7917\pm0.0333$ & 0.703 \\
 & Order+entropy & $0.7868\pm0.0347$ & 0.690 \\
 & Fixed subset, $k=2$ & $0.6933\pm0.0583$ & 0.667 \\
\midrule
\multirow{5}{*}{MSU/ORNL}
 & Full fusion & $0.3924\pm0.0077$ & 1.000 \\
 & DQN ($\lambda=0.5$) & $0.3769\pm0.0118$ & 0.754 \\
 & Order+entropy & $0.3859\pm0.0131$ & 0.787 \\
 & Fixed subset, $k=2$ & $0.3931\pm0.0093$ & 0.400 \\
 & Fixed subset, $k=3$ & $\mathbf{0.4011\pm0.0101}$ & 0.600 \\
\bottomrule
\end{tabular}
\end{table}

Classifiers built for one subset could change this answer, so for (Q2) we repeated the study on three group disjoint splits with validation selected tree models (Table~\ref{tab:hardening}). On TU Delft the DQN stays stable and fixed order entropy remains slightly ahead, but the larger shift comes from the classifier, since HGB reading only the network modality, which validation picked on all three splits, reaches $0.8673\pm0.0272$ at one third of the cost. We report HGB because it was the strongest tree family on MSU/ORNL, and on TU Delft RF and XGBoost stay within 0.013 of it at every $k$. Two and three modality HGB reach 0.878 and 0.908 at higher cost, the chosen pair differing across splits, so the earlier gain over a fixed two source mask reflects the shared classifier and does not show that sequential acquisition is the best way to deploy. MSU/ORNL gives a clearer answer still, as two and three modality HGB beat the DQN by 0.067 and 0.076 Macro-F1 at costs of 0.40 and 0.60 against 0.728. Reading all five modalities adds little (0.4402), and validation chose relays R1, R3, and R4 for the three modality model on every split, so when a small and stable subset carries most of the signal, classifier choice matters more than the acquisition policy.

\begin{table}[t]
\centering
\tabstyle
\caption{Hardening results over partition seeds 42, 123, and 456.}
\label{tab:hardening}
\begin{tabular}{llcc}
\toprule
\hdrrow
\midrule
\multirow{5}{*}{TU Delft}
 & DQN & $0.8045\pm0.0165$ & 0.706 \\
 & Order+entropy & $0.8095\pm0.0113$ & 0.732 \\
 & HGB, $k=1$ & $0.8673\pm0.0272$ & 0.333 \\
 & HGB, $k=2$ & $0.8779\pm0.0292$ & 0.667 \\
 & HGB, $k=3$ & $\mathbf{0.9077\pm0.0466}$ & 1.000 \\
\midrule
\multirow{5}{*}{MSU/ORNL}
 & DQN & $0.3632\pm0.0268$ & 0.728 \\
 & Order+entropy & $0.3800\pm0.0264$ & 0.751 \\
 & HGB, $k=2$ & $0.4299\pm0.0088$ & 0.400 \\
 & HGB, $k=3$ & $0.4389\pm0.0069$ & 0.600 \\
 & HGB, $k=5$ & $\mathbf{0.4402\pm0.0061}$ & 1.000 \\
\bottomrule
\end{tabular}
\end{table}

Table~\ref{tab:hardening} still sets a neural sequential policy against tree models, so Table~\ref{tab:hgbseq} removes the classifier family confound by running entropy stopping over the HGB models themselves. On TU Delft, calibrated sequential HGB beats static HGB by 0.031 at lower cost (0.571 against 0.778), reading the network modality first on every split and using 1.7 of the three sources on average. Its mean even exceeds the three source HGB of Table~\ref{tab:hardening} (0.9077), which per sample selection allows when subset models err on different samples, although the margin lies within the spread across splits and we did not trace its source. With validation cost capped at two thirds, it beats the best static subset within that cap on all three splits by 0.039 to 0.100, although the cap holds on validation cost only and test cost can drift above it (0.724 on partition 123). On MSU/ORNL, static and sequential HGB are tied in Macro-F1, sequential HGB at cost 0.861 saves nothing over the three relay HGB of Table~\ref{tab:hardening} at cost 0.60, and under a cap of 0.4 it trails the static subset by 0.023. These results answer (Q2), because the shared backbone DQN result does not carry over to a practical deployment frontier, yet a simple sequential rule over strong per subset classifiers still pays off on TU Delft.

\begin{table}[t]
\centering
\tabstyle
\caption{Static VS sequential HGB over partition seeds 42, 123, and 456. Static HGB takes the subset with the best Macro-F1 on the second validation half, without a cost cap.}
\label{tab:hgbseq}
\begin{tabular}{llcc}
\toprule
\hdrrow
\midrule
\multirow{3}{*}{TU Delft}
 & Static HGB & $0.8972\pm0.0580$ & 0.778 \\
 & Sequential HGB, raw & $0.9254\pm0.0443$ & 0.520 \\
 & Sequential HGB, calibrated & $\mathbf{0.9282\pm0.0378}$ & 0.571 \\
\midrule
\multirow{3}{*}{MSU/ORNL}
 & Static HGB & $0.4376\pm0.0077$ & 0.933 \\
 & Sequential HGB, raw & $0.4395\pm0.0057$ & 0.879 \\
 & Sequential HGB, calibrated & $\mathbf{0.4400\pm0.0066}$ & 0.861 \\
\bottomrule
\end{tabular}
\end{table}

\begin{figure}[t]
\centering
\includegraphics[width=\columnwidth]{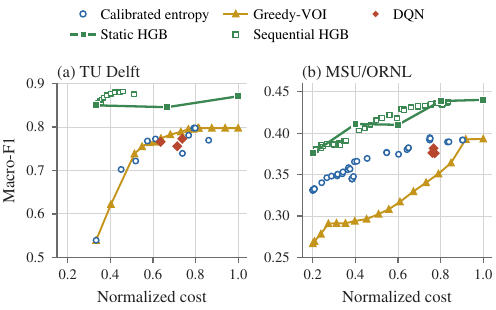}
\caption{Macro-F1 against normalized cost on the primary split, with the backbone of initialization seed 42 and the HGB models of partition 42.}
\label{fig:frontier}
\end{figure}

\begin{figure}[t]
\centering
\includegraphics[width=\columnwidth]{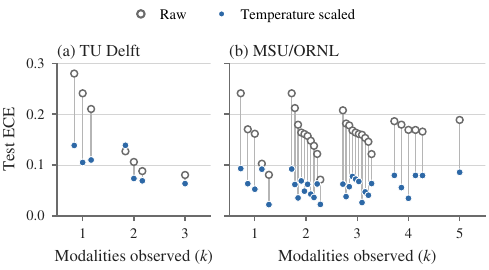}
\caption{Mask-wise test ECE before and after temperature scaling (seed 42).}
\label{fig:calib}
\end{figure}

Returning to the shared backbone, temperature scaling per mask strengthens the simple sequential baseline for (Q1) by lowering mean test ECE on both datasets (Table~\ref{tab:calrob}), on all 31 MSU/ORNL masks, and on six of the seven TU Delft masks (Fig.~\ref{fig:calib}), the exception being the physical and resource pair, which rises slightly from 0.127 to 0.139. With calibrated probabilities, entropy stopping and Greedy-VOI cover large parts of the shared backbone Pareto frontier (Fig.~\ref{fig:frontier}), where the entropy points are settings chosen on validation, the DQN points are the three $\lambda$ values, and the HGB curves come from the partition 42 models on the same test set. Static HGB lies above every shared backbone policy across the whole cost range, and sequential HGB fills the gaps between static subsets, scoring on TU Delft above every static subset in the figure at under half the cost.

Across three $\lambda$ values on each of the initialization seeds 42, 123, and 456, no TU Delft DQN point is Pareto optimal for seeds 42 or 123, whereas all three points of seed 456 are, and on MSU/ORNL none of the nine DQN points is Pareto optimal against the calibrated and simple controls, so we read the frontier as a whole and name no single best stopping policy. The group bootstrap for initialization seed 42 gives a DQN minus entropy difference of $-0.0135$ on TU Delft (95\% interval $[-0.0419,0.0142]$), whose sign differs from the ten seed paired result although both intervals contain zero, and $-0.0150$ on MSU/ORNL (interval $[-0.0300,-0.0039]$), which favors the simpler entropy rule.

\begin{table}[t]
\centering
\tabstyle
\caption{Mask-wise calibration of the initialization seed 42 backbone and first modality corruption of its DQN.}
\label{tab:calrob}
\begin{tabular}{lccc}
\toprule
Dataset & ECE raw$\rightarrow$cal. & $\Delta$F1 zero & $\Delta$F1 noise \\
\midrule
TU Delft & $0.162\rightarrow0.100$ & $-0.371$ & $-0.107$ \\
MSU/ORNL & $0.163\rightarrow0.059$ & $-0.033$ & $-0.134$ \\
\bottomrule
\end{tabular}
\end{table}

\begin{figure}[htb]
\centering
\includegraphics[width=\columnwidth]{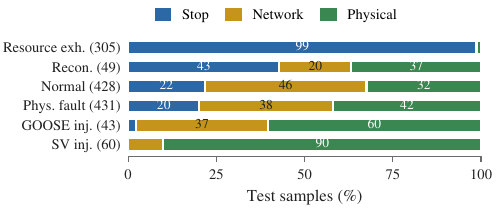}
\caption{DQN decision after resource telemetry on TU Delft (seed 42).}
\label{fig:branch}
\end{figure}

For (Q3), deterministic policies start from the same empty state, so adaptivity shows in what the policy does after its first observation. On TU Delft, the initialization seed 42 DQN first acquires resource telemetry and then stops at once for 98.7\% of resource exhaustion samples but for only 2.3\% of GOOSE injection samples, sends 90.0\% of SV injection samples on to physical evidence, and gives normal, fault, and reconnaissance cases their own mix of STOP and the two remaining sources (Fig.~\ref{fig:branch}).

That branching is also an attack surface, which we probe by zeroing the standardized features of the first acquired modality or adding Gaussian noise $\mathcal{N}(0,2^2)$ to them (Table~\ref{tab:calrob}). On TU Delft, zeroing drops Macro-F1 from 0.7656 to 0.3951 and noise drops it to 0.6585, and both raise the mean cost as the policy keeps asking for more evidence. On MSU/ORNL, noise hurts more than zeroing and lowers the mean cost from 0.761 to 0.620 because the policy stops earlier, which is consistent with overconfident posteriors on noisy inputs, although we did not measure this directly. These stress tests are not adversarial guarantees, and because only the DQN was corrupted, the drops mix policy and classifier effects. They still answer (Q3) by showing that evidence selection should be tested under corrupted early observations, since spoofing the first source a deterministic policy reads could trigger STOP before the revealing modality is read, as the earlier stopping under noise on MSU/ORNL hints.

\section{Discussion and Limitations}
\label{sec:discussion}
Taken together, the results separate two questions that are easy to conflate, namely whether sequential acquisition helps and whether it needs a learned policy. The shared backbone protocol isolates the policy from classifier differences, and under it the DQN clearly beats a fixed two source mask on TU Delft. Deployment is not tied to one backbone, though, and with strong per subset classifiers a single network modality or a compact HGB subset beats the learned policy at much lower cost, while stopping over those classifiers does better still on TU Delft. Adaptivity also does not require RL, as calibrated entropy stopping and Greedy-VOI show with far less training machinery. On action spaces this small, a learned value function has to earn its place through a reproducible gain on the frontier of Macro-F1 against cost, and in our experiments it rarely did, so the main engineering lesson is that sequential acquisition helped most as a plain stopping rule on top of strong classifiers.

\textbf{Limitations.} Equal modality cost ignores measured bytes, latency, energy, storage I/O, and operator effort. We did not train a DQN over the tree posteriors, so the RL comparison still rests on the neural backbone. Three $\lambda$ values per seed give the DQN a sparse frontier against the dense threshold sweeps of the stopping rules, which favors the sweeps when Pareto points are counted. We use only two datasets, the hardening analysis has three split replications, and rare TU Delft classes remain small. Resource exhaustion onsets are coarse, and our threat model is known class diagnosis rather than open set detection. Finally, we did not reproduce feature level acquisition systems such as JAFA \cite{Shim2018}, GSM \cite{Li2021}, or ACO \cite{Valancius2024}, since a fair comparison at the modality level would need its own implementation and tuning study.

\section{Conclusion}
\label{sec:conclusion}
This paper asked how much evidence a power system diagnoser should collect before it decides, and whether a learned policy should make that choice. Our experiments do not show that a more sophisticated acquisition policy is always better. The DQN learns sample dependent paths and reads less evidence than full fusion, yet calibrated stopping rules stay competitive with it, and single modality HGB on TU Delft and three modality HGB on MSU/ORNL beat it with higher Macro-F1 at lower normalized cost. Entropy stopping over those classifiers reaches the highest mean Macro-F1 we report on TU Delft while reading 57\% of the sources, and adds nothing on MSU/ORNL. Selective evidence access is therefore worth studying, but practitioners should first test classifier strength, calibration, static subsets, and simple sequential rules before they adopt a learned policy, and should test any selective scheme against corrupted or forged early evidence. Natural next steps are measured, unequal source costs and stop inducing attacks on the first acquired source.

\balance

\end{document}